\documentclass[12pt]{article}
\usepackage{amssymb}
\usepackage{graphicx}
\usepackage{authblk}
\begin{document}

\title{Energy transfer in elastic collisions between electrons and trapped ions}

\author{{\L}ukasz K{\l}osowski\thanks{lklos@fizyka.umk.pl} }
\author{ Mariusz Piwi\'nski} 
\affil{Institute of Physics, Faculty of Physics, Astronomy and Informatics, Nicolaus Copernicus University in Toru\'n, Grudzi{\c a}dzka 5, 87-100 Toru\'n, Poland}

\maketitle

\begin{abstract}
Heating of trapped ion clouds by interactions with free electrons crossing the trapping potential was observed. 
A model describing such process was proposed and discussed. 
The presented approach predicts two effects: pushing and heating of the ions' ensemble by electrons. 
The former was found to be too weak for observation, however the latter was investigated experimentally.
For comparison between experimental results and theoretical predictions, molecular dynamics simulations for various ion ensembles in various temperatures were performed to determine  dependence between ion cloud geometry and its temperature. 
A heating rate coefficient was defined and determined together with temperatures of ion clouds bombarded with electrons. 
Good correlation between the two quantities was found, which agrees with the proposed model.
\end{abstract}

\section{Introduction}
Elastic electron-ion collisions are important phenomena  for low temperature plasmas,
atmospheric, astrophysical research, etc., as the interaction between charged particles is
the most fundamental for modeling ionized gases. They ave been widely studied
both experimentally and theoretically 
(see the review by M\"uller \cite{mill08} and references therein, as well as some later works \cite{jung13,kang18}).

In the experiment on electron impact ionization for trapping purposes \cite{klos18}, some effects  were observed \cite{klos17b}, which can be interpreted as a consequence of  electron scattering on trapped ions \cite{mcqu15}.

As the ions (calcium in this case) are produced in electron-atom collisions, trapped, and optically cooled down \cite{klos17a}, it is possible to observe directly the reactions of ion ensembles to presence of other particles, such as electrons. 
In the experiment it was observed, that captured ion ensemble is temporarily expanded in space when it is bombarded with electrons,  
while the total number of trapped ions remains  unchanged.
Such effect can be interpreted as a consequence of kinetic energy transfer from electron to ion in elastic collisions \cite{pohl05}. 

Electron impact on a trapped ion can lead to one of several consequences:
\begin{itemize}
\item Elastic collision, which does not cause  ion loss from the trap directly. However, if the kinetic energy accumulated by the ion in multiple collisions is sufficient to pass the trap's potential barrier, an ion can be lost. 
\item Inelastic collision, where ion changes its internal quantum state. Similarly to elastic process, this does not affect the number of trapped ions.
\item Recombination of ion: Ca$^++e^-\longrightarrow $Ca.
\item Further ionization: Ca$^++e^-\longrightarrow $Ca$^{2+}+2e^-$, Ca$^++e^-\longrightarrow $Ca$^{3+}+3e^-$, etc.
\end{itemize}
The two last effects can reduce the number of ions inside the trap. 
As the elastic collision is described with long-range Coulomb potential, the first process appears to be dominating over other three and ionization or recombination are less probable.

\section{Theoretical model for distribution of ions in trapping potential}
\label{temp}
To build up the theoretical model, two extreme cases should be considered: 
an ensemble of ions at high temperature allowing to neglect Coulomb repulsion for analysis of their geometry and low temperature ensemble forming a Coulomb crystal. 

Effective potential of the linear Paul trap can be well approximated as anisotropic, harmonic one, where depth of the potential remains the same in both directions perpendicular to the  trap's main axis:
\begin{equation}
V(x,y,x)=\frac M2\left(\omega_{xy}\left(x^2+y^2\right)+\omega_zz^2\right) ,
\label{1}
\end{equation}
where $M$ is the ion mass and $\omega$ are frequencies of ion motion.

If there is a single ion trapped in such potential, its motions in various degrees of freedom are independent harmonic oscillations. 
This way, the ion's energy can be divided into three  parts $E_{x}$, $E_{y}$ and $E_{z}$.
The amplitudes of oscillations can be expressed as:
\begin{equation}
x_0=\frac{\sqrt{\frac{2E_x}M}}{\omega_x} ,
\end{equation}
and similar for $y_0$ and $z_0$.

The probability distribution of finding an ion in a certain position $x$ in a classical oscillator model is given with expression:
\begin{equation}
f(x)=\frac{1}{\pi\sqrt{x_0^2-x^2}}=\frac{1}{\pi\sqrt{\frac{2E_x}{M\omega_x^2}-x^2}}
\label{distrib}
\end{equation}
and analogous for $y$ and $z$.

As the image exposure time in the experimental conditions (of the order of 1 second) is much longer than ion oscillation period (tens of microseconds), the intensity distribution of ion image should be described with function from equation (\ref{distrib}).

If a large ensemble of ions is trapped and their temperature is high enough to neglect ion-ion repulsion energy, then the ensemble can be treated as a set of independent particles. 
The Coulomb interactions allow however for exchange of kinetic energies between ions, which leads to thermal equilibrium.
Energies of individual ions yield Boltzmann distribution, so the image intensity function should be averaged over the possible ion energies, which can be expressed as convolution of equation (\ref{distrib}) and Boltzmann distribution at given temperature $T$:
\begin{equation}
F(x)=\int_{\frac{M\omega_x^2x^2}{2}}^{\infty}\frac{\exp\left(-\frac{E_x}{kT}\right)}{\pi kT\sqrt{\frac{2E_x}{M\omega_x^2}-x^2}}{\rm d} E_x =
\sqrt{\frac{M\omega_x^2}{2\pi kT}}\exp\left(-\frac{M\omega_x^2x^2}{2kT}\right)
\label{splot}
\end{equation}
resulting in a Gaussian distribution of the ion cloud, whose width $\sigma_{\infty x}$ depends on temperature:
\begin{equation}
\sigma_{\infty x}=\sqrt{\frac{kT}{M\omega_x^2}}=\sqrt{\frac{E_x}{M\omega_x^2}} .
\label{sx}
\end{equation}
Analogous expressions for $\sigma_{\infty y}$ and $\sigma_{\infty z}$ widths can be written.

The other extreme case is temperature close to 0 K,  where ions form a Coulomb crystal. 
Its size depends on the number of ions $N$ and depth of the trapping potential. 
In harmonic trap's field, the number density of trapped ions should not depend on $N$, and the aspect ratio of a crystal should be independent from $N$ in given trap settings.
This way, the size of crystallized ion ensemble $\sigma_{0x}$ (and analogous $\sigma_{0y}$ and $\sigma_{0z}$) can be found as:
\begin{equation}
\sigma_{0x}=A_x\cdot N^{\frac13},
\end{equation}
where $A_x$ (and analogous $A_y$ and $A_z$) is a coefficient depending on trapping potential. 

The sizes of ion cloud $\sigma_x(T)$, $\sigma_y(T)$, $\sigma_z(T)$ should be expressed with functions, which tend asymptotically to $\sigma_{\infty x}(T)$, $\sigma_{\infty y}(T)$, $\sigma_{\infty z}(T)$ at $T\rightarrow \infty $ and and their values at $T=0$ should be $\sigma_{0x}$,   $\sigma_{0y}$ and $\sigma_{0z}$.
Expressions defining such functions should be found numerically.

\subsection{Temperature-cloud size correlation}
\label{montecarlo}
According to the previous paragraph, equation (\ref{splot}) is valid at higher temperatures. 
At lower temperatures the repulsive forces between ions would result in additional broadening of the cloud, which in general can be difficult to predict analytically. 
To determine impact of the effect, Monte Carlo simulations of molecular dynamics of ion ensembles were performed.

The simulation applies classical equations of motion for a set of particles of given mass and charge.
The ions are placed in an effective trapping potential from equation (\ref{1}). 
Their initial conditions are chosen randomly: positions in 1mm cube and velocities with 1000~$\frac{\rm m}{\rm s}$ maximum absolute value. 
The simulation was performed using midpoint method with viscosity term allowing for cooling the ion ensemble. 
As the viscosity model is non-physical for particles in vacuum, the simulated dynamics of ions is not reliable, however the final state of the ensemble is close to the laboratory situation where the ions are Doppler-cooled. 
The viscosity is achieved by reducing all ion velocities by some constant factor (slightly below 1) in each step of simulation.
This way, after sufficient number of steps, the simulated kinetic energy of ions is reduced and the ions reach their equilibrium positions.

To allow higher temperatures of ion ensembles, a maximum velocity of ions parameter $v_{max}$ is introduced in calculations. 
The viscosity is ''switched off'' for ions of the speed below $v_{max}$.

The simulation is continued until the system reaches stable geometry (not necessary a crystal). 
As the ion set fluctuates, 1000 last steps are used for calculations of statistical properties of the ensembles. 
Such approach allows to reduce statistical spread of the numerical results. 

\begin{figure*}[t]
\begin{center}
\includegraphics[width=\textwidth]{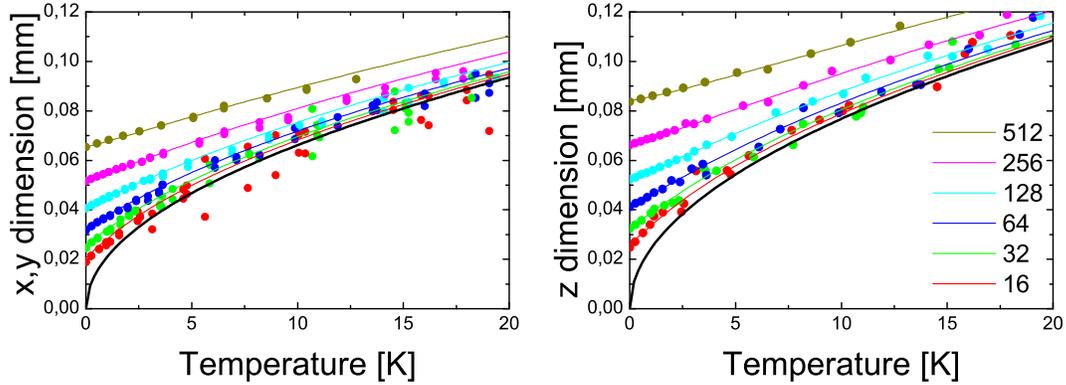}
\caption{Dimensions of the ion ensembles at various ion numbers and temperatures. 
The black solid line is a cloud size expected in a single-ion/hot-ensemble model (equation (\ref{sx}) and analogous). 
The points are numerically determined, where  sizes defined as a single standard deviation of ion positions averaged over 1000 steps of calculations. The temperatures are calculated from mean kinetic energy of the ions,  averaged the same way. 
The color lines refer to equation (\ref{sssx}) fitted to numerical data for various numbers of ions. 
The ion numbers are marked with colors as labeled in the image.
 } 
\label{temperatura}
\end{center}
\end{figure*}

The simulations were performed for various numbers of ions and various final energies (temperatures) of ion ensembles. 
Example results are presented in  figure \ref{temperatura}. 
They were obtained for trapping conditions of: $M=40$, $\omega_{x,y}=9.75\cdot10^5$ s$^{-1}$, $\omega_z=8.4\cdot10^5$ s$^{-1}$ and $N$ from 16 to 512 ions.
\begin{figure*}[t]
\begin{center}
\includegraphics[width=\textwidth]{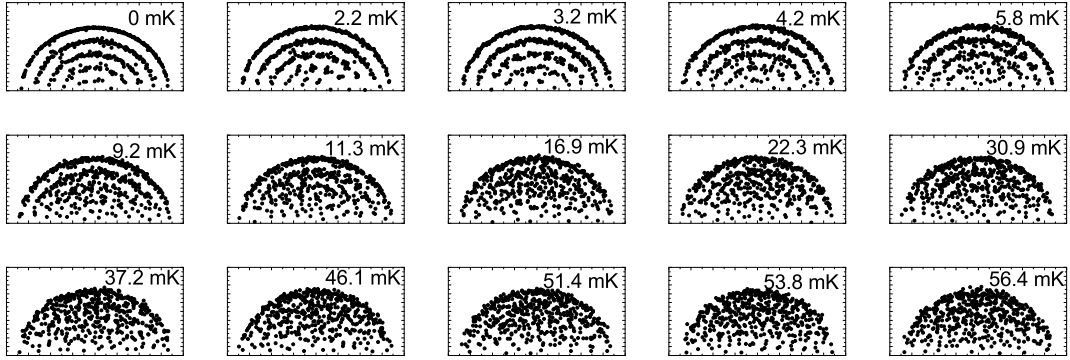}
\caption{Numerically derived clouds of 512 ions at sub-kelvin temperatures.
The figures are projections of the ensembles along the azimuthal coordinate. The vertical axis is the distance of an ion from the trap's main axis.
All panels are 400 $\mu$m wide and 200 $\mu$m high.} 
\label{crystals}
\end{center}
\end{figure*}
The obtained data show, that the sizes of ion ensembles can be very well approximated with functions:
\begin{equation}
\sigma_x(N,T)=\sqrt[\kappa]{\sigma_{\infty x}^\kappa+\sigma_{0x}^\kappa}=\left({\left(\frac{kT}{M\omega_x^2}\right)^{\frac\kappa2}+\left(A\cdot N^{\frac13}\right)^\kappa}\right)^{\frac1\kappa} ,
\label{sssx}
\end{equation}
where $\kappa$ is exponent depending on trapping potential. 
Analogous expressions can be written for $\sigma_y$ and $\sigma_z$.

Analysis of the numerical simulation results (figure \ref{temperatura}), provides values:
$\kappa=2.26$, $A=8.2\mu$m for $x,y$ directions and
$\kappa=2.50$, $A=10.6\mu$m for $z$ direction.

Additionally, the ensembles of 512 ions were simulated at sub-kelvin temperatures. 
The results are presented in figure \ref{crystals} as projections of ion ensembles along azimuthal coordinate to distinguish between  
crystallized and non-crystallized ones. 
The main conclusion from these images is that for the lowest temperatures, the size of observed ensemble does not change with temperature. 
This way, Coulomb crystallization is not necessary here to estimate the number of ions inside the trap.

\section{Theoretical model of electron-ion scattering}
\begin{figure*}[t]
\begin{center}
\includegraphics[width=\textwidth]{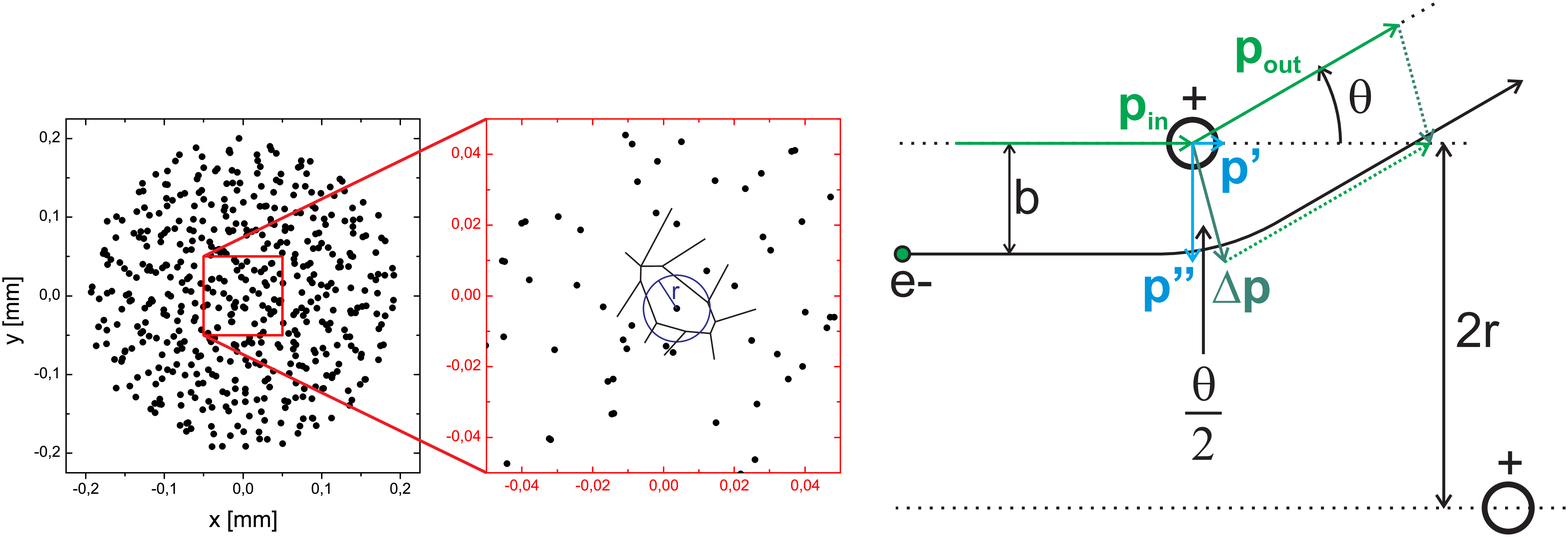}
\caption{Definitions of parameters used in the Rutherford-Coulomb scattering model. 
The left panel represents a projection of ion ensemble (obtained numerically) along the electron beam direction. 
The central part is enlarged.
An example of a cell -- an area, where the chosen ion is the closest one to the electron trajectory -- is presented. 
The loop of the radius $r$ has the same area as the cell. 
The right panel defines the classical scattering parameter $b$, the scattering angle $\theta$, linear momenta of incoming and outgoing electron ${\bf p_{in}}$ and ${\bf p_{out}}$ respectively, the momentum transferred to the ion ${\bf\Delta p}$  and its longitudinal and transversal parts 
${\bf p'}$ and ${\bf p''}$. } 
\label{model}
\end{center}
\end{figure*}
To understand the kinetic energy and linear momentum transfer from electrons to ions, a~collision model should be proposed. 
The one used in this work is a Rutherford model \cite{ruth11} of elastic collision with attractive Coulomb potential.

In case of the electron-ion scattering,  Coulomb potential is dominating and thus, only such term is taken into account. 
As the classical and quantum scattering models for such potential provide the same results, the classical one is used here for simplicity. 
In the experiments, where scattering potential differs slightly from Coulomb, some differences between cross sections derived from Rutherford model and more complex ones can be observed \cite{hube94,brot02,srig96a,srig96b,srig96c}. 
They are however seen in the regions far from $\theta=0$ maximum of scattering amplitude, so the differences are significantly reduced in averaging of the cross sections.

Velocities of ions at sub-kelvin temperatures are below  $10{\rm\frac ms}$. 
On the other hand, velocity of 10 eV electron is of the order of $10^{6}{\rm\frac ms}$. 
Also electron mass is 4 orders of magnitude lower than ion's.
Thus, the ions can be treated as stationary scattering centers for electrons. 
Therefore the experiment can be modeled as an electron penetrating through a multi-center, scattering potential of Coulomb character.

The ion ensemble is relatively sparse, with number density $n$ of the order of $10^6{\rm mm}^{-3}$.
If the ion cloud is projected on a plane, then a surface density $\varsigma$ can be defined and expressed as:
\begin{equation}
\varsigma= N^{\frac13}n^{\frac23} ,
\label{varsigma}
\end{equation}
and a single ion ''covers'' a cell of area of $S=\frac1\varsigma$ in the electron beam cross section, which is equivalent of a circle of radius $r$ (see figure \ref{model}):
\begin{equation}
r= \pi^{-\frac12}N^{-\frac16}n^{-\frac13} .
\label{r}
\end{equation}
The probability, that electron would pass two or more ions in close proximity is rather low here, so one can assume single scattering --  electron interaction with just one ion, the closest one to electron trajectory, can be taken into account.

Classical scattering parameter $b$ should be then lower than the radius $r$.
In classical Rutherford-Coulomb model \cite{ruth11}, the scattering angle is given with expression:
\begin{equation}
\theta=2\arctan\frac{2e^2}{4\pi\varepsilon_0Eb} ,
\end{equation}
where $E$ is electron energy.
The linear momentum transfer is then:
\begin{equation}
\left| \Delta {\bf p} \right| = 2\sqrt{2Em}\cdot{\sin\frac\theta2} ,
\end{equation}
where $m$ is the electron mass. 
The momentum transfer is divided between longitudinal $p'$ and transversal $p''$ parts:
\begin{equation}
p' = \sqrt{2Em}\cdot(1-\cos\theta) ,
\end{equation}
\begin{equation}
p'' = \sqrt{2Em}\cdot{\sin\theta} .
\end{equation}
According to the momentum conservation principle, the same amount of momentum (with opposite sign) is gained by the ion. 
To determine an average transfer of the linear momentum, one should integrate only the longitudinal part, as the transversal parts are equally probable for all the directions and result in zero average value. 
The integration should be performed over the $\frac1\varsigma$ area. 
For simplicity it is approximated with a disk (the contribution of the edges of the integration region is very low, so it should not significantly affect the result):
\begin{equation}
\left<p\right> = \frac{1}{\pi r^2}\int_0^{2\pi}\int_0^r\sqrt{2Em}\cdot(1-\cos\theta)b{\rm d}b{\rm d}\varphi
 = \sqrt{8Em}\frac{\ln\left(\left(\frac{4\pi\varepsilon_0E}{2e^2}\right)^2r^2+1\right)}{\left(\frac{4\pi\varepsilon_0E}{2e^2}\right)^2r^2} .
\label{p}
\end{equation}

The momentum transfer results also in an energy transfer. 
The magnitude of the  transferred energy depends not only on the electron initial state, but also on the ion motion before collision. 
It can be shown, that if average expected motion of ion is zero, then the expected value of energy transfer is equal to the one from stationary ion: 
\begin{equation}
 \Delta E = \left(2\sqrt{2Em}\cdot{\sin\frac\theta2}\right)^2\frac1{2M}= \frac{4Em}{M}{\sin^2\frac\theta2} ,
\end{equation}
which can be averaged over the $\pi r^2$ area:
\begin{equation}
\left< \Delta E \right>=\frac{1}{\pi r^2}\int_0^{2\pi}\int_0^r \frac{4Em}{M}{\sin^2\frac\theta2}b{\rm d}b{\rm d}\varphi
=\frac{4 Em}{M}\frac{\ln\left(\left(\frac{4\pi\varepsilon_0E}{2e^2}\right)^2r^2+1\right)}{\left(\frac{4\pi\varepsilon_0E}{2e^2}\right)^2r^2} .
\label{e}
\end{equation}
In equations (\ref{p}) and (\ref{e}) a characteristic dimensionless parameter $\eta$ can be defined:
\begin{equation}
\eta=\frac{4\pi\varepsilon_0rE}{2e^2}=\frac{E}{2\frac{e^2}{4\pi\varepsilon_0r}} .
\end{equation}
At typical experimental conditions $E=100$eV, $N=1000$ and $n=10^6$mm$^{-3}$, the parameter has value of $\eta\approx10^3$, which allows to neglect ''1'' in the logarithm in equations (\ref{p}) and (\ref{e}).
They can be then written in simpler forms:
\begin{equation}
\left<p\right> = \sqrt{32Em}\frac{\ln\eta}{\eta^2}=\sqrt{\frac{16e^2m}{\pi\varepsilon_0r} }\cdot\frac{\ln\eta}{\eta^{\frac32}} ,
\end{equation}
\begin{equation}
\left< \Delta E \right>=\frac{8 Em}{M}\frac{\ln\eta}{\eta^2}=\frac{ {4e^2} m}{\pi\varepsilon_0rM}\cdot\frac{\ln\eta}{\eta} ,
\label{de}
\end{equation}
both functions being of descending character with $\eta$ in the considered range of energies. 

\subsection{Collective effect on ion ensemble}
When electron beam is used, multiple scattering events for single electrons cumulate to observable effects of pushing and heating of ion ensembles.
If $I$ is current of the beam, then  frequency of electron passages $\nu$ would be: 
\begin{equation}
\nu=\frac Ie .
\end{equation}
The effective force $F$ pushing the ion cloud can be expressed as:
\begin{equation}
F=\left<p\right>\nu .
\label{f}
\end{equation}
A force of the same value acting on ensemble of $N$ ions can be written as
\begin{equation}
F=NeE_e ,
\end{equation}
where $E_e$ is a value of some electric field. Thus the electric field equivalent to the pushing effect can be calculated:
\begin{equation}
E_e=\frac{\left<p\right>I}{Ne^2} .
\end{equation}
At typical conditions of electronic beam of the order of $1\mu$A and other parameters  
\mbox{$E=100$eV}, $N=1000$, $n=10^6$mm$^{-3}$, one obtains $E_e\approx10^{-3}\frac{\rm V}{\rm m}$.
Such value of the electric field is much lower than possible fluctuations of the potentials of the trap electrodes, so the pushing effect is too low to be noticed in present experimental conditions.

Similarly to equation \ref{f}, the power $P$ heating up the ion ensemble can be calculated as:
\begin{equation}
P=\left<E\right>\nu .
\end{equation}
The power $P_1$ absorbed by a single ion is then:
\begin{equation}
P_1=\frac{\left<E\right>I}{Ne} .
\label{p1}
\end{equation}
In the same experimental conditions as above, this leads to a value of $P_1$ of the order of $10\frac{\rm meV}{\rm s}$.

Additionally, one can define a heating rate $\gamma$ of ions:
\begin{equation}
\gamma=\frac{{\rm d}T}{{\rm d}t} ,
\end{equation}
where $T$ is ions' temperature and $t$ is the time. The heating parameter can be then found as:
\begin{equation}
\gamma=\frac23\frac{P_1}{k_B} ,
\label{gama}
\end{equation}
where $k_B$ is Boltzmann constant. 
In the considered experimental conditions, $\gamma$ is of the order of $100\frac{\rm K}{\rm s}$. 
Thus the heating effect can be observed experimentally as an increase in temperature of trapped ions resulting in expansion of the ion cloud when bombarded with electrons. 

If one assumes, that in experimental conditions the electron beam is broader than ion ensemble and its current density $j$ is approximately uniform, then the effective electric current $I$ can be expressed as:
\begin{equation}
I=j\cdot S=j\cdot N^{\frac23}n^{-\frac23} ,
\end{equation}
where  $S$ is the area covered by ion ensemble.
The heating rate, by combining equations (\ref{varsigma}), (\ref{r}), (\ref{de}),  (\ref{p1}) and (\ref{gama}), can be  expressed in explicit way:
\begin{equation}
\gamma=\frac{4mje^3}{3k_BM
{\pi^2\varepsilon_0^2E}
}
\left(\ln\frac{2\sqrt\pi\varepsilon_0n^{-\frac13}E}{e^2}-\frac16\ln N\right) ,
\label{wyrazenie}
\end{equation}
which provides relatively simple relation between $\gamma$ coefficient and number of ions trapped~$N$.

\section{Experimental results}
The apparatus used in experiment was described in our previous works \cite{klos18,klos17a,klos17,klos18a}. 
It consists of a linear, segmented Paul trap inside a vacuum chamber. 
The calcium ions were produced in electron-atom collisions inside the trap. 
The electron beam was produced by a low-energy, pulsed gun and the atomic beam came from resistively heated oven. 
\begin{figure*}[t]
\begin{center}
\includegraphics[width=\textwidth]{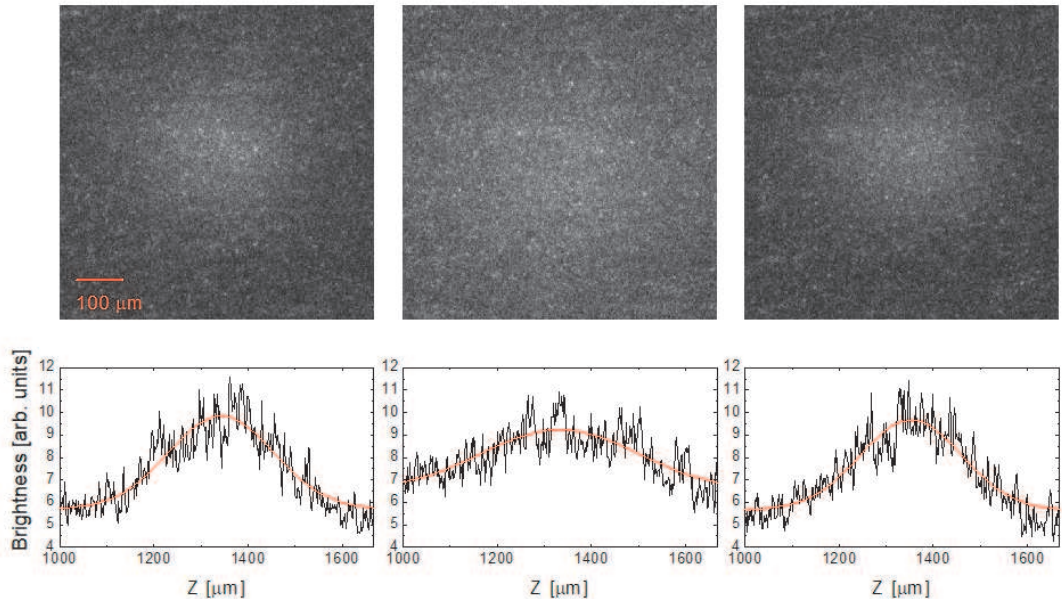}
\caption{Example images of ion clouds for 180 eV electron beam: (a) before interaction with electrons, (b) during the interaction, (c) after the interaction. 
There are no significant differences between images from panels (a) and (c), however the cloud in the panel (b) seems significantly enlarged. 
The bottom panels represent cross sections through the images along $Z$ axis (horizontal) in the central parts of the ion cloud. 
The red lines are Gaussian fitting of the cross sections.} 
\label{zdjecia}
\end{center}
\end{figure*}

The trap is driven with $1\div 4$ MHz AC voltage with several volts DC between central and outer segments of the trap. 
Trap settings were chosen to be away from any nonlinear resonances \cite{klos18a}.
In this particular case,  1 MHz frequency of the trap AC voltage was used. Potential depths were $\omega_{x,y}=9.75\cdot10^5$ s$^{-1}$ and $\omega_z=8.4\cdot10^5$ s$^{-1}$ (the same as used in calculations in section \ref{montecarlo}).

Doppler cooling system \cite{klos17a} consisted of two stabilized diode lasers of 397 and 866 nm. 
Imaging system was an image-intensified camera equipped with optical filter allowing for detection of 397 nm fluorescence of ions.
Number and temperature of ions can be estimated by analysis of the image of the ensemble.

Number of ions can be found in two different ways, depending on their temperature and trap settings: 
If the ions are well cooled down to form a Coulomb crystal \cite{wine87}, then the ion number determines the crystal's size. 
This way, by measurement of crystal's diameter, one can find the number with low uncertainty.

If the ions are not crystallized, which is observed in electron-impact-heated ensembles, then the number can be estimated from size and brightness of the ion cloud, which provides much higher uncertainty in comparison to the first method. 
The temperature of non-crystallized ion ensemble can be found by determining its size and using equation (\ref{sssx}).

In the experiment, the ion clouds were captured in the same conditions (temperature of atomic oven, electron energy, beam intensity, electronic pulse duration and trap settings) before every electron-ion scattering shot. 
This way, the expected number of ions in the trapped ensemble should be approximately the same before every experimental run.

In the next step, a 60-second waiting period was taken to allow cooling down the calcium oven and reaching dynamic equilibrium of the  
ion ensemble.
A single, 3-second electronic pulse was introduced with various energies from 20 to 200 eV. 
The ions were dropped from the trap after another 60 second waiting and a new ensemble of ions could be prepared for the next shot.
The further analysis included the images of ion clouds before the electron pulse, after 2 seconds of bombardment and several seconds after the  electronic irradiation. 
 
The electron pulse was limited to 3 seconds for several reasons: 
\begin{itemize}
\item The ion ensemble expands during the bombardment, which causes change in the value of $\gamma$ coefficient. 
To avoid more complex analysis of the ion heating dynamics, the expansion was reduced by short pulse duration and $\gamma$ was only slightly changed.
\item If the ion ensemble expanded to the size comparable with the trap, some of the ions could be lost by leaving the trapping potential well or by hitting one of the electrodes of the trap. Also larger ensembles would excess the size of the region observed by the camera.
\item Hotter ion ensembles would detune from the Doppler cooling laser frequency too far to provide sufficient fluorescence for imaging.
\end{itemize}

The ion cloud images are recorded every 1 second with CCD camera. 
This way, the the electron pulse duration covers at least two complete frames of the recorded images. 
These frames can be compared to the ones recorded before and after the electron pulse and this way the number of ions trapped $N$ and their temperature $T$  with electron beam can be estimated.

Density of current $j$ of electrons can be found by  simulating numerically their spatial distributions and 
measuring the currents collected by trap electrodes in the way discussed in our previous paper \cite{klos18}.
Subsequently, the heating rate $\gamma$ can be determined using equation (\ref{wyrazenie}).

\subsection{Results}
An example of ion ensembles is presented in figure \ref{zdjecia} for electron energy of 180 eV.
For the images, width $w$ and height $h$ of the clouds were determined as a width of Gaussian profile fitted to image cross sections. 
Before bombarding the ions, the sizes were \mbox{$w=(218\pm8)$ $\mu$m}, $h=(192\pm8)$  $\mu$m. 
According to equation (\ref{sssx}), one can estimate the number of trapped ions, assuming low temperature, to be $N=1190\pm140$.
During the interaction, the dimensions were expanded to $w=(348\pm20)$  $\mu$m, $h=(382\pm15)$  $\mu$m, which corresponds to temperature of about $(375\pm57)$ K. 
After the interaction, the cloud shrank back to its original sizes $w=(212\pm10)$  $\mu$m, $h=(228\pm8)$  $\mu$m. The number of ions can be estimated to be $N=1360\pm150$, which is not significantly different from the one before electron bombardment.

\begin{figure*}[t]
\begin{center}
\includegraphics[width=.7\textwidth]{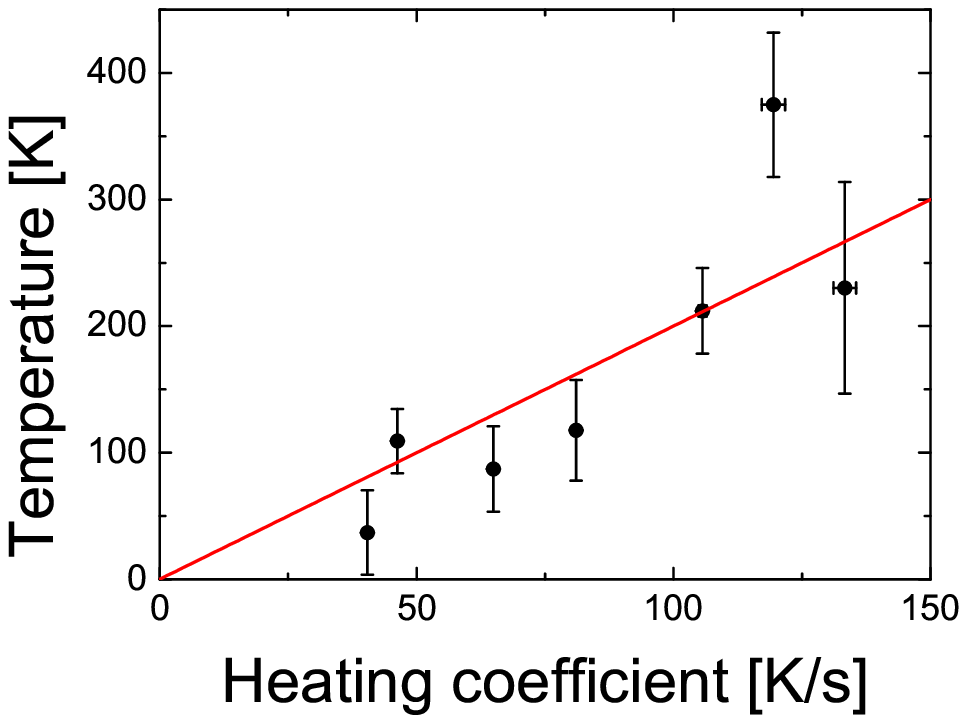}
\caption{Results of measurement of the ion temperature versus heating coefficient $\gamma$. 
The temperatures are read after 2 seconds of bombardment of ions with electrons. 
Besides relatively large error bars for temperatures, there is a good correlation between both measured quantities.
The red line is expected $T=\gamma\cdot2{\rm s}$ function graph.} 
\label{wynik}
\end{center}
\end{figure*}

The experiment  was repeated for various electron beam energies from 20 to 200 eV.
As the effect below 90 eV was to weak to provide conclusive results, only the energies above 90 eV are further discussed. 
The experimental results are collected in table \ref{tabela} and presented in figure \ref{wynik}.
\begin{table*}[t]
\begin{center}
\caption{The values of heating coefficient $\gamma$ and temperature of ion ensemble after 2~seconds of bombardment with electrons.}
\label{tabela}
\begin{tabular}{cccc}
\hline \hline
Electron	&	electron beam								&	estimated 									&	temperature of 	\\
energy [eV]	&	intensity $[\frac{\rm \mu A}{\rm mm^2}]$	&	$\gamma \left[\frac{\rm K}{\rm s}\right]$	&	ion ensemble [K]	\\
\hline
90	& $	1.13	\pm	.02	$ & $	40.4	\pm	0.3	$ & $	37\pm	34	$ \\
100	& $	1.42	\pm	.02	$ & $	46.3	\pm	0.3	$ & $	109	\pm	26	$ \\
120	& $	2.40	\pm	.02	$ & $	65.0	\pm	0.5	$ & $	87	\pm	34	$ \\
140	& $	3.47	\pm	.02	$ & $	81.0	\pm	0.5	$ & $	118	\pm	40	$ \\
160	& $	5.06	\pm	.02	$ & $	105.7	\pm	0.9	$ & $	212	\pm	34	$ \\
180	& $	6.37	\pm	.12	$ & $	119.4	\pm	2.3	$ & $	375	\pm	57	$ \\
200	& $	7.86	\pm	.12	$ & $	133.3	\pm	2.3	$ & $	230	\pm	84	$ \\
\hline \hline
\end{tabular}
\end{center}
\end{table*} 
In general, one can conclude there is a correlation between the temperature reached in 2 seconds and the heating coefficient. 
One can expect the $T$ value should be approximately twice the $\gamma$ value which was confirmed experimentally.

The relatively large error bars for temperature determination are mainly due to image fluctuations resulting in uncertainties of observed cloud sizes. 
As the derivatives of functions from equation (\ref{sssx}) are low, the uncertainties of temperatures obtained experimentally are relatively high and such an issue can be difficult to overcome.

\section{Summary and conclusions}
The heating effect for ions inside a linear Paul trap by their interaction with free electrons was observed and  described using a simple, classical scattering model. 
The proposed model predicts also an associated effect of electronic pressure acting on ions, however it is too weak to be observed experimentally. 

The proposed approach, besides being relatively simple, very well predicts magnitude of the effect of collisional heating, which  agrees with experimental results.  

Described model can be widely applied for analysing various interaction between ions and relatively light charged particles, such as phenomena observed in plasma, atmospheric or astrophysical research.

It can be also useful for estimation of the limits of application of electronic beams in cold atoms/ions experiments.

\section*{Acknowledgements}
This work has been supported by the National Science Centre, Poland,  project no.~2014/13/B/ST2/02684.


\end{document}